\documentclass[conference,a4paper]{IEEEtran}
\IEEEoverridecommandlockouts

\usepackage[left=1.62cm, right=1.62cm,top=1.9cm,bottom=4.4cm]{geometry}

\usepackage{silence}
\usepackage{cite}

\usepackage{amsmath,amssymb,amsthm,fixmath}

\usepackage{color,colortbl}
\usepackage[table]{xcolor}
\definecolor{rowgray}{gray}{0.92}

\usepackage[inline]{enumitem}

\usepackage{siunitx}
\DeclareSIUnit{\dBm}{dBm}

\usepackage{graphicx}
\usepackage[labelformat=simple]{subcaption}

\usepackage{epstopdf}
\usepackage{cuted}
\usepackage{multirow,booktabs}
\usepackage{diagbox}
\usepackage{makecell}
\usepackage{hhline}
\usepackage{array} 
\newcolumntype{x}{!{\vrule width 2px}}
\newcolumntype{y}{!{\vrule width 1.5px}}

\usepackage{optidef}

\usepackage{nohyperref} 

\usepackage[shortcuts,acronym,automake]{glossaries}
\makeglossaries
\newacronym{6g}{6G}{Sixth Generation}
\newacronym{awgn}{AWGN}{additive white Gaussian noise}
\newacronym{bcd}{BCD}{block coordinate descent}
\newacronym{bec}{BEC}{binary erasure channel}
\newacronym{bler}{BLER}{block error rate}
\newacronym{blec}{BLEC}{block erasure channel}
\newacronym{bri}{BRI}{biregular irreducible}
\newacronym{cdf}{CDF}{cumulative distribution function}
\newacronym{clt}{CLT}{central limit theorem}
\newacronym{cp}{CP}{control plane}
\newacronym{crc}{CRC}{cyclic redundancy check}
\newacronym{csi}{CSI}{channel state information}
\newacronym{csit}{CSIT}{channel state information at transmitter}
\newacronym{dft-s-ofdm}{DFT-s-OFDM}{Discrete Fourier Transform-spread-OFDM}
\newacronym{dl}{DL}{deep learning}
\newacronym{fbl}{FBL}{finite blocklength}
\newacronym{gan}{GAN}{generative adversarial network}
\newacronym{harq}{HARQ}{hybrid automatic repeat request}
\newacronym{jscc}{JSCC}{joint source-channel coding}
\newacronym{kl}{KL}{Kullback-Leibler}
\newacronym{ibl}{IBL}{infinite blocklength}
\newacronym{ik}{IK}{incremental knowledge}
\newacronym{ir}{IR}{incremental redundancy}
\newacronym{ldpc}{LDPC}{low-density parity-check}
\newacronym{lfp}{LFP}{leakage-failure probability}
\newacronym{ls}{LS}{least squares}
\newacronym{mac}{MAC}{medium access control}
\newacronym{mcs}{MCS}{modulation and coding scheme}
\newacronym{mrc}{MRC}{maximal ratio combining}
\newacronym{mimo}{MIMO}{multi-input multi-output}
\newacronym{ml}{ML}{maximum likelihood}
\newacronym{mm}{MM}{Minorize-Maximization}
\newacronym{noma}{NOMA}{non-orthogonal multi-access}
\newacronym{nom}{NOM}{non-orthogonal multiplexing}
\newacronym{ofdm}{OFDM}{orthogonal frequency-division multiplexing}
\newacronym{ofdma}{OFDMA}{orthogonal frequency-division multiple access}
\newacronym{oma}{OMA}{orthogonal multiple access}
\newacronym{papr}{PAPR}{Peak-to-Average Power Ratio}
\newacronym{pdf}{PDF}{probability density function}
\newacronym{per}{PER}{packet error rate}
\newacronym{phy}{PHY}{physical}
\newacronym{pld}{PLD}{physical layer deception}
\newacronym{pls}{PLS}{physical layer security}
\newacronym{prb}{PRB}{physical resource block}
\newacronym{psk}{PSK}{phase-shift keying}
\newacronym{qos}{QoS}{quality of service}
\newacronym{semcom}{SemCom}{semantic communication}
\newacronym{sic}{SIC}{successive interference cancellation}
\newacronym{sinr}{SINR}{signal-to-interference-and-noise ratio}
\newacronym{snr}{SNR}{signal-to-noise ratio}
\newacronym{tdma}{TDMA}{time-division multiple access}
\newacronym{up}{UP}{user plane}
\newacronym{urllc}{URLLC}{ultra-reliable low-latency communication}
\newacronym{vae}{VAE}{variational autoencoder}
\newacronym{xurllc}{xURLLC}{next-generation URLLC}

\usepackage{tcolorbox}
\tcbuselibrary{many}

\usepackage[norelsize,linesnumbered,ruled]{algorithm2e}
\SetKwRepeat{Do}{do}{while}
\makeatletter
\newcommand{\removelatexerror} {\let\@latex@error\@gobble}
\makeatother

\usepackage{flushend}

\usepackage[normalem]{ulem}

\usepackage[textsize=tiny,colorinlistoftodos]{todonotes}
\makeatletter
\define@key{todonotes}{bh}[]{
	\setkeys{todonotes}{author=\textbf{Bin}, color=lime!30}}%
\define@key{todonotes}{yz}[]{
	\setkeys{todonotes}{author=\textbf{Yao}, color=blue!30}}%
\define@key{todonotes}{rs}[]{
	\setkeys{todonotes}{author=\textbf{Rafael}, color=green!30}}%
\makeatother

\tikzstyle{note}=[rectangle, minimum width=3cm, draw = none, fill = none, minimum width = 1.5cm, anchor=center, align=left]
\tikzstyle{block}=[rectangle, draw, line width=1pt, fill = none, minimum width = 1cm, minimum height = 0.75cm, anchor=center, inner sep = 0.5mm, align=center]
\tikzstyle{arrow} = [thick,->,>=stealth]

\newif\ifreviewmode
\reviewmodefalse

\ifreviewmode
\else
  \renewcommand{\todo}[1]{} 
\fi

\begin{document}

\title{Generative Semantic HARQ:\\Latent-Space Text Retransmission and Combining}

\author{
	\IEEEauthorblockN{
		Bin~Han\IEEEauthorrefmark{1}
		, Yulin~Hu\IEEEauthorrefmark{2}, 
        and~Hans~D.~Schotten\IEEEauthorrefmark{1}\IEEEauthorrefmark{3} 
	}
	\IEEEauthorblockA{
		\IEEEauthorrefmark{1}RPTU University Kaiserslautern-Landau\qquad
		\IEEEauthorrefmark{2}Wuhan University
		\\
		\IEEEauthorrefmark{3}German Research Center for Artificial Intelligence (DFKI GmbH)
	}
}

\maketitle

\begin{abstract}
	Semantic communication conveys meaning rather than raw bits, but reliability at the semantic level remains an open challenge. We propose a semantic-level \ac{harq} framework for text communication, in which a Transformer-\ac{vae} codec operates as a lightweight overlay on the conventional protocol stack. The stochastic encoder inherently generates diverse latent representations across retransmissions---providing \ac{ik} from a single model without dedicated protocol design. On the receiver side, a soft quality estimator triggers retransmissions and a quality-aware combiner merges the received latent vectors within a consistent latent space. We systematically benchmark six semantic quality metrics and four soft combining strategies under hybrid semantic distortion that mixes systematic bias with additive noise. The results suggest combining Weighted-Average or MRC-Inspired combining with self-consistency-based \ac{harq} triggering for the best performance.
\end{abstract}


\glsresetall

\section{Introduction}\label{sec:intro}

Driven by the rising demand for post-Shannon communication systems, \ac{semcom} has emerged as a promising paradigm that transmits meaning rather than raw bits, achieving higher communication efficiency especially under bandwidth-constrained conditions~\cite{NRA+2026contemporary}. Deep learning-based \ac{semcom} systems, such as DeepSC~\cite{XQLJ2021deep} for text and Deep \ac{jscc}~\cite{BKG2019deep} for images, have demonstrated significant gains over conventional separate source-channel coding by jointly optimizing the encoding and decoding processes end-to-end across the semantic and physical layers.

While the majority of \ac{semcom} research has focused on visual data (images, videos, point clouds) for their high compression potential, text-oriented \ac{semcom} remains comparatively under-explored. This is despite the growing importance of textual semantics as \begin{enumerate*}[label=\emph{\roman*)}]
	\item the interface layer between conventional communication systems and emerging token-based multimedia communications, where compact textual descriptions convey visual or audio content,
	\item a key modality for cross-modal translation tasks such as text-to-image and text-to-speech generation, and
	\item a native representation for the semantic plane in AI-RAN architectures.
\end{enumerate*}

To enhance the reliability of \ac{semcom}, several works have extended \ac{harq} mechanisms to the semantic level. Jiang et al.~\cite{JWJ+2022deep} proposed the first end-to-end semantic \ac{harq} for sentence transmission (SCHARQ), employing chase combining and incremental redundancy in the latent space. Zhou et al.~\cite{ZLZ+2022adaptive} introduced \ac{ik}-based \ac{harq} (IK-HARQ) with adaptive bit-rate control. Beyond text, semantic \ac{harq} has been investigated for image transmission~\cite{ZWX+2025semantic, ZHCY2024scan}, feature transmission~\cite{hwx+2026semharq, WLMF2025spiking}, video conferencing~\cite{JWJ+2023wireless}, and cooperative perception~\cite{SLYJ+2026semantic}. However, three challenges remain. First, unlike conventional \ac{harq} where \ac{crc} on the \ac{mac} layer triggers retransmissions and soft combining operates on the \ac{phy} layer, semantic-level \ac{harq} must integrate error detection and correction within the semantic layer itself, requiring new quality assessment methods beyond bit-level checks. Second, all existing semantic \ac{harq} approaches employ \emph{deterministic} encoders, so retransmissions carry identical or structurally constrained representations. Third, while individual works propose ad-hoc solutions for quality estimation and combining, no systematic comparison of these methods exists.

In this work, we propose a generative semantic \ac{harq} framework for text communication that addresses these gaps. Our main contributions are:
\begin{enumerate*}[label=\emph{\roman*)}]
	\item We design a Transformer-\ac{vae} based semantic encoder-decoder that serves as a \emph{lightweight overlay} on top of conventional communication systems. The stochastic nature of the \ac{vae} naturally provides diverse latent representations across retransmissions, enabling \ac{ik} without explicit protocol design.
	\item We systematically compare six semantic quality metrics---ranging from encoder uncertainty and latent-space distance to round-trip consistency checks---as retransmission triggers, benchmarking their effectiveness under the same experimental conditions.
	\item We systematically compare four semantic combining strategies---from softmax-weighted averaging to iterative refinement approaches---against the baseline of chase combining, evaluating their ability to exploit multiple received latent representations.
\end{enumerate*}

The remainder of this paper is organized as follows. Section~\ref{sec:approach} presents the proposed framework, including the system architecture, quality estimation methods, and combining strategies. Section~\ref{sec:evaluation} provides the performance evaluation. Section~\ref{sec:conclusion} concludes the paper.

\section{Proposed Approach}\label{sec:approach}

\subsection{System Architecture}\label{sec:architecture}

\begin{figure*}[t]
	\centering
	\includegraphics[width=.8\linewidth]{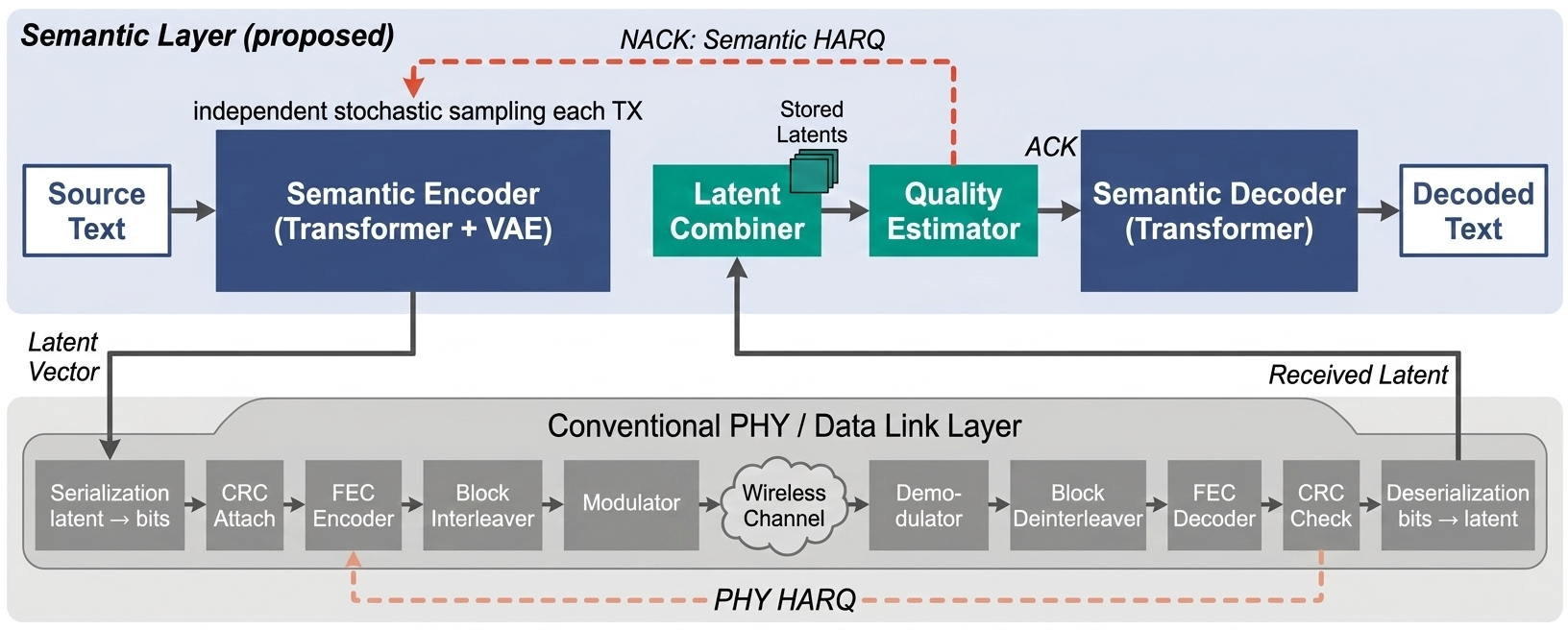}
	\caption{The proposed framework of semantic-level \ac{harq} as a lightweight overlay on conventional communication systems.}
	\label{fig:architecture}
\end{figure*}

The proposed system architecture is illustrated in Fig.~\ref{fig:architecture}. A semantic codec pair operates as a lightweight overlay on top of the conventional communication stack. The semantic encoder maps a source sentence $\mathbf{s}$ into a compact latent vector $\mathbf{z} \in \mathbb{R}^D$, where $D$ is the latent dimension, which is transmitted through the conventional \ac{phy} chain (channel coding, modulation, etc.) and received as $\tilde{\mathbf{z}}$. On the receiver side, a semantic decoder reconstructs the sentence as $\hat{\mathbf{s}} = f_\text{dec}(\tilde{\mathbf{z}})$.

At the lower layers, conventional channel coding combined with \ac{crc}-based \ac{harq} can generally provide sufficient integrity of the data packets. The propagation of bit errors upward into the semantic layer is therefore very rare and practically negligible. Nevertheless, the received latent $\tilde{\mathbf{z}}$ may still differ from the transmitted $\mathbf{z}$ due to distortions that originate within or above the semantic layer itself, motivating a dedicated semantic-level \ac{harq} mechanism. We identify two categories of such distortion that may coexist. The first is \emph{unbiased, additive distortion}, which is approximately zero-mean and i.i.d.\ across transmissions. Sources include quantization and rounding errors from fixed-point or reduced-precision inference on edge devices, residual channel noise that survives \ac{phy}-layer error correction but still perturbs the latent, and intentionally injected noise for purposes such as differential privacy. The second category is \emph{biased distortion}, which introduces a systematic shift in the latent space and is \emph{not} i.i.d.\ across transmissions. Sources include encoder-decoder misalignment due to asynchronous model updates (e.g., one side is updated while the other retains an older version), deployment asymmetry where the encoder and decoder run on heterogeneous hardware with different model variants (e.g., full-precision vs.\ distilled), and domain shift between the training distribution and the deployment context. Biased distortion is particularly challenging because it persists across retransmissions and cannot be mitigated by simple averaging---i.e., chase combining of identical latent representations. Effective semantic \ac{harq} therefore requires \emph{\ac{ik}}: each retransmission must carry a distinct representation of the same semantic content, so that the combiner can exploit diversity rather than merely averaging out noise. This fundamentally requires a semantic encoder whose output varies across retransmissions of the same source, which motivates the generative codec design presented next.

The semantic \ac{harq} operates as follows. After each transmission round $k$, a quality estimator computes a score $q_k \in [0,1]$ for the received latent $\tilde{\mathbf{z}}_k$. If $q_k$ falls below a predefined threshold $q_\text{th}$, a NACK triggers retransmission. After $K$ rounds (or upon ACK), the receiver combines all received latents $\{\tilde{\mathbf{z}}_1, \ldots, \tilde{\mathbf{z}}_K\}$ into $\hat{\mathbf{z}}$ and decodes the final output $\hat{\mathbf{s}} = f_\text{dec}(\hat{\mathbf{z}})$. The design of the codec, quality estimator, and combiner are detailed in the following subsections.

\subsection{Generative Semantic Codec}\label{sec:codec}

As established above, effective semantic \ac{harq} relies on \ac{ik}, i.e., each retransmission must carry a structurally distinct representation of the same semantic content. A straightforward realization would employ multiple, separately trained encoders---analogous to \ac{ir} on lower-layer \ac{harq}---but this multiplies model storage and maintenance cost on the transmitter side. We instead adopt a Transformer-\ac{vae} architecture whose single stochastic encoder inherently produces diverse latent representations via random sampling. Because every sample is drawn from the same learned latent space~$\mathbb{R}^D$, the semantic-to-latent mapping remains consistent across retransmissions and all received vectors $\{\tilde{\mathbf{z}}_1,\ldots,\tilde{\mathbf{z}}_K\}$ reside in the same space. Combining therefore reduces to a fixed-dimensional vector operation whose output can be directly fed to the \emph{same} decoder, regardless of the number of retransmissions~$K$. This avoids the dimensionality growth that would arise from concatenating heterogeneous representations, which would require $K$-specific decoders on the receiver and impose a hard cap on the maximum number of retransmissions, since each encoder variant must be provisioned in advance. The consistent latent space also simplifies quality estimation, as all received vectors are directly comparable.

The encoder processes a source sentence $\mathbf{s}$ through tokenization, embedding with positional encoding, and $L$ layers of multi-head self-attention with feed-forward networks. The resulting contextualized representations are mean-pooled and projected into the \ac{vae} latent space:
\begin{equation}\label{eq:vae_encode}
	(\boldsymbol{\mu}, \log\boldsymbol{\sigma}^2) = f_\text{enc}(\mathbf{s}), \quad \mathbf{z} = \boldsymbol{\mu} + \boldsymbol{\sigma} \odot \boldsymbol{\epsilon}, \quad \boldsymbol{\epsilon} \sim \mathcal{N}(\mathbf{0}, \mathbf{I}),
\end{equation}
where $\odot$ denotes the element-wise product. The reparameterization enables gradient-based training while maintaining stochasticity. Each retransmission draws a fresh noise sample $\boldsymbol{\epsilon}$, producing a distinct latent $\mathbf{z}_k$ that encodes the same semantic content from a different region of the latent space%
.

The decoder projects the received latent $\tilde{\mathbf{z}}$ into a cross-attention memory and an input injection embedding, then generates the output sentence autoregressively via $L$ Transformer decoder layers with causal masking. The procedures are detailed in Algorithms~\ref{alg:encode} and~\ref{alg:decode}. In both algorithms, $\textsc{Tok}(\cdot)$ and $\textsc{Detok}(\cdot)$ denote tokenization and detokenization with respect to vocabulary $\mathcal{V}$; $\textsc{Emb}(\cdot)$ is the shared token embedding layer; $\textsc{PE}(\cdot)$ adds sinusoidal positional encoding; $\textsc{TfEnc}(\cdot)$ and $\textsc{TfDec}(\cdot)$ denote the $L$-layer Transformer encoder and decoder blocks; $\textsc{Pool}(\cdot)$ performs mean pooling over non-padded positions; and $\langle\text{S}\rangle$, $\langle\text{E}\rangle$, $\langle\text{P}\rangle$ are the start-of-sequence, end-of-sequence, and padding tokens. The binary padding mask $\mathbf{M}_\text{pad}$ indicates non-padded positions.

The model is trained end-to-end by minimizing a combined reconstruction and \ac{kl} divergence loss over an \ac{awgn} channel:
\begin{equation}\label{eq:loss}
	\mathcal{L} = \mathcal{L}_\text{recon} + \beta \cdot \mathcal{L}_\text{KL},
\end{equation}
where $\mathcal{L}_\text{recon}$ is the cross-entropy between the decoder output and the target tokens, and the \ac{kl} term with free-bits regularization is
\begin{equation}\label{eq:kl}
	\mathcal{L}_\text{KL} = -\frac{1}{2}\sum_{d=1}^{D} \max\!\big(1 + \log\sigma_d^2 - \mu_d^2 - \sigma_d^2,\; \lambda_\text{free}\big).
\end{equation}
The weighting factor $\beta$ follows a linear annealing schedule to prevent posterior collapse. During each training step, the \ac{snr} is randomly sampled from a predefined set $\mathcal{S}$ to ensure robustness across channel conditions. The training and validation procedures are summarized in Algorithm~\ref{alg:train}, where $\textsc{Reparam}(\cdot)$ denotes the reparameterization trick in~\eqref{eq:vae_encode}, $\textsc{AWGN}(\cdot)$ adds channel noise at the given \ac{snr}, and $\textsc{ClipGrad}(\cdot)$ clips gradients by norm.

\begin{algorithm}[!htpb]
	\scriptsize
	\DontPrintSemicolon
	\caption{Semantic Encoding (Transmitter)}\label{alg:encode}
	\KwIn{Source sentence $\mathbf{s}$, vocabulary $\mathcal{V}$}
	\KwOut{Latent vector $\mathbf{z} \in \mathbb{R}^D$}
	$\mathbf{t} \gets \textsc{Tok}(\mathbf{s}, \mathcal{V})$\tcp*{$[\langle\text{S}\rangle, w_1, \ldots, w_n, \langle\text{E}\rangle, \langle\text{P}\rangle, \ldots]$}
	$\mathbf{X} \gets \textsc{Emb}(\mathbf{t}) \cdot \sqrt{d_\text{model}} + \textsc{PE}(|\mathbf{t}|)$\;
	$\mathbf{H} \gets \textsc{TfEnc}(\mathbf{X}, \mathbf{M}_\text{pad})$\tcp*{$L$ self-attention layers}
	$\mathbf{h} \gets \textsc{Pool}(\mathbf{H}, \mathbf{M}_\text{pad})$\;
	$\boldsymbol{\mu} \gets \mathbf{W}_\mu \mathbf{h} + \mathbf{b}_\mu$, \quad $\log\boldsymbol{\sigma}^2 \gets \mathbf{W}_\sigma \mathbf{h} + \mathbf{b}_\sigma$\;
	$\boldsymbol{\epsilon} \sim \mathcal{N}(\mathbf{0}, \mathbf{I})$\;
	$\mathbf{z} \gets \boldsymbol{\mu} + \exp(0.5 \cdot \log\boldsymbol{\sigma}^2) \odot \boldsymbol{\epsilon}$\;
	\Return{$\mathbf{z}$}
\end{algorithm}
\vspace{-3mm}

\begin{algorithm}[!htpb]
	\scriptsize
	\DontPrintSemicolon
	\caption{Semantic Decoding (Receiver)}\label{alg:decode}
	\KwIn{Received latent $\tilde{\mathbf{z}}$, vocabulary $\mathcal{V}$}
	\KwOut{Reconstructed sentence $\hat{\mathbf{s}}$}
	$\mathbf{m} \gets \mathbf{W}_\text{proj}\, \tilde{\mathbf{z}}$\tcp*{cross-attention memory}
	$\mathbf{z}_\text{emb} \gets \mathbf{W}_\text{emb}\, \tilde{\mathbf{z}}$\tcp*{input injection}
	$\text{out} \gets [\langle\text{S}\rangle]$\;
	\For{$t = 1, 2, \ldots, t_\text{max}$}{
		$\mathbf{Y} \gets \textsc{Emb}(\text{out}) \cdot \sqrt{d_\text{model}} + \mathbf{z}_\text{emb} + \textsc{PE}(t)$\;
		$\mathbf{D} \gets \textsc{TfDec}(\mathbf{Y}, \mathbf{m}, \mathbf{M}_\text{causal})$\;
		$w_\text{next} \gets \arg\max\, \mathbf{W}_\text{out}\, \mathbf{D}[t]$\;
		$\text{out} \gets \text{out} \,\|\, [w_\text{next}]$\;
		\lIf{$w_\text{next} = \langle\text{E}\rangle$}{\textbf{break}}
	}
	$\hat{\mathbf{s}} \gets \textsc{Detok}(\text{out}, \mathcal{V})$\;
	\Return{$\hat{\mathbf{s}}$}
\end{algorithm}

\begin{algorithm}[!htpb]
	\scriptsize
	\DontPrintSemicolon
	\caption{Training and Validation}\label{alg:train}
	\KwIn{$\mathcal{D}_\text{train}$, $\mathcal{D}_\text{val}$, SNR set $\mathcal{S}$, max epochs $N_e$, patience $P$}
	$\theta \gets \textsc{InitParams}()$; \quad $\phi \gets \textsc{AdamW}(\theta)$; \quad $\eta \gets \textsc{ReduceLR}(\phi)$\;
	$m^* \gets -\infty$; \quad $c \gets 0$\tcp*{best metric; stall counter}
	\For{$e = 1, \ldots, N_e$}{
		$\beta \gets \textsc{Anneal}(e)$\tcp*{KL weight warm-up}
		\ForEach{$(\mathbf{s}, \mathbf{t}_\text{in}, \mathbf{t}_\text{out}) \in \mathcal{D}_\text{train}$}{
			$\gamma \gets \textsc{Sample}(\mathcal{S})$\tcp*{random SNR}
			$(\boldsymbol{\mu}, \log\boldsymbol{\sigma}^2) \gets f_\text{enc}(\mathbf{s})$\;
			$\mathbf{z} \gets \textsc{Reparam}(\boldsymbol{\mu}, \boldsymbol{\sigma})$\;
			$\tilde{\mathbf{z}} \gets \textsc{AWGN}(\mathbf{z}, \gamma)$\;
			$\hat{\mathbf{t}} \gets f_\text{dec}(\tilde{\mathbf{z}}, \mathbf{t}_\text{in})$\;
			$\mathcal{L} \gets \textsc{CE}(\hat{\mathbf{t}}, \mathbf{t}_\text{out}) + \beta \cdot \mathcal{L}_\text{KL}$\tcp*{Eqs.~\eqref{eq:loss}--\eqref{eq:kl}}
			$\nabla \gets \textsc{ClipGrad}(\nabla_\theta \mathcal{L})$; \quad $\theta \gets \phi.\textsc{Step}(\nabla)$\;
		}
		$m \gets \textsc{Eval}(\mathcal{D}_\text{val})$\tcp*{BLEU-4, similarity}
		$\eta \gets \textsc{UpdateLR}(\eta, m)$\tcp*{reduce on plateau}
		\eIf{$m > m^*$}{$m^* \gets m$; \quad $c \gets 0$; \quad save $\theta$\;}{$c \gets c + 1$\;}
		\If{$c \geq P$}{\textbf{break}\tcp*{early stopping}}
	}
\end{algorithm}
\vspace{-3mm}

\subsection{Semantic Quality Estimation}\label{sec:quality}
In conventional \ac{harq}, retransmission is triggered by a binary \ac{crc} check that unambiguously detects bit errors. At the semantic level, no such clear-cut criterion exists: semantic meaning is inherently tolerant of minor variations, and an overly strict detector would forfeit the very flexibility that makes \ac{semcom} attractive. The challenge is therefore to design a \emph{soft} quality metric that distinguishes genuine semantic degradation from benign variation. All estimation is performed at the receiver. We consider six metrics that produce a scalar score $q \in [0,1]$ (higher is better), summarized in Table~\ref{tab:qe_methods}. The methods differ in their computational requirements at the receiver. Metrics~A and~D require re-encoding the decoded sentence to obtain the encoder statistics $\boldsymbol{\mu}$ and $\boldsymbol{\sigma}$, but involve no additional decoding step. Metric~C requires only the decoder, running $N$ independent decoding passes to measure output consistency. Metrics~B, E, and~F perform a full round-trip: decode the received latent, re-encode the result to obtain $\boldsymbol{\mu}'$, and then decode $\boldsymbol{\mu}'$ again to assess consistency at the text level (Metrics~E,~F) or latent level (Metric~B). In Metric~B, $\delta(\boldsymbol{\mu}, \boldsymbol{\mu}') = \max\!\big(0,\, 1 - \|\boldsymbol{\mu} - \boldsymbol{\mu}'\|_2 / \sqrt{D}\big)$ is a normalized distance penalty; $\epsilon = 10^{-8}$ in Metric~D prevents division by zero.

\vspace{-1mm}
\begin{table}[!htpb]
	\centering
	\caption{Semantic quality metrics.}
	\label{tab:qe_methods}
	\renewcommand{\arraystretch}{1.3}
	\begin{tabular}{@{}cllc@{}}
		\toprule[2px]
		& \textbf{Metric} && \textbf{RX overhead} \\
		\midrule[1px]
		\rowcolor{rowgray}
		A & VAE Uncertainty: &$q = \max\!\big(0,\, 1 - \frac{1}{D}\sum_{d} \sigma_d\big)$ & Re-enc. \\
		B & Self-Consistency: &$q = \cos(\boldsymbol{\mu}, \boldsymbol{\mu}') \cdot \delta(\boldsymbol{\mu}, \boldsymbol{\mu}')$ & Re-enc.+dec. \\
		\rowcolor{rowgray}
		C & Decoder Entropy: &$q = \frac{1}{N}\sum_{n} \mathbb{1}[\hat{\mathbf{s}}^{(n)} = \hat{\mathbf{s}}^{(1)}]$ & $N{\times}$dec. \\
		D & Latent Distance: &$q = \max\!\big(0,\, 1 - \frac{\|\tilde{\mathbf{z}} - \boldsymbol{\mu}\|}{\|\boldsymbol{\mu}\| + \epsilon}\big)$ & Re-enc. \\
		\rowcolor{rowgray}
		E & Text BLEU: &$q = \text{BLEU-1}(\hat{s}_1, \hat{s}_2)$ & Re-enc.+dec. \\
		F & Text Similarity: &$q = J(\hat{s}_1, \hat{s}_2)$ & Re-enc.+dec. \\
		\bottomrule[2px]
	\end{tabular}
\end{table}

\vspace{-3mm}
\subsection{Semantic Combining}\label{sec:combining}

Because the \ac{vae} encoder---as discussed in Section~\ref{sec:codec}---maps every retransmission into the same latent space~$\mathbb{R}^D$, the combiner's task reduces to merging $K$ vectors that share a common coordinate system into a single estimate $\hat{\mathbf{z}} \in \mathbb{R}^D$, which is then passed to the unchanged decoder. This fixed-dimensional formulation allows us to draw on classical combining principles from \ac{phy}-layer diversity reception and adapt them to the semantic domain.

We consider four quality-aware strategies, summarized in Table~\ref{tab:combining_methods}. Method~A applies softmax-normalized weights derived from the quality scores, providing a smooth weighting that retains every received transmission. Method~B can be viewed as an extreme variant that zeros out all weights except for the highest-quality transmission. Method~C, inspired by classical \ac{mrc}, strikes a compromise between the two: it still incorporates every transmission, but its quadratic weighting rule favors high-quality receptions far more aggressively than the exponential softmax of Method~A, since on $q_k\in[0,1]$ the quadratic mapping compresses low scores much more strongly. Method~D takes an iterative approach, starting from the highest-quality reception and progressively blending in the remaining vectors.

\begin{table}[!htpb]
	\centering
	\caption{Semantic combining methods.}
	\label{tab:combining_methods}
	\renewcommand{\arraystretch}{1.3}
	\begin{tabular}{@{}cllc@{}}
		\toprule[2px]
		& \textbf{Method} & & \textbf{Complexity} \\
		\midrule[1px]
		\rowcolor{rowgray}
		A & Weighted-Avg.: &$\hat{\mathbf{z}} = \sum_k w_k \tilde{\mathbf{z}}_k$, $w_k = \frac{e^{q_k}}{\sum_j e^{q_j}}$ & $\mathcal{O}(KD)$ \\
		B & Best-Only: &$\hat{\mathbf{z}} = \tilde{\mathbf{z}}_{k^*}$, $k^* = \arg\max_k q_k$ & $\mathcal{O}(K)$ \\
		\rowcolor{rowgray}
		C & MRC-Inspired: &$\hat{\mathbf{z}} = \frac{\sum_k q_k^2 \tilde{\mathbf{z}}_k}{\sum_k q_k^2}$ & $\mathcal{O}(KD)$ \\
		& &$\hat{\mathbf{z}}^{(0)}\!=\!\tilde{\mathbf{z}}_{k^*}$,&\\
		D & Iterative: &$\hat{\mathbf{z}}^{(t+1)}\!=\!\alpha\hat{\mathbf{z}}^{(t)}+(1\!-\!\alpha)\tilde{\mathbf{z}}_k$,& $\mathcal{O}(KD)$ \\
		& &$\alpha\!=\!\frac{q^{(t)}}{q^{(t)}+q_k}$  &\\
		\bottomrule[2px]
	\end{tabular}
\end{table}

\section{Performance Evaluation}\label{sec:evaluation}

\subsection{Evaluation Setup}\label{sec:setup}

We evaluate on the MS-COCO 2014 Train/Val annotations dataset~\cite{COCO2014}, comprising $400~172$ English sentences (5--50 words each) with a vocabulary of $|\mathcal{V}|=10~000$ tokens, split into $320~137$ training, $40~017$ validation, and $40~018$ test sentences (80/10/10\%). The model and training hyperparameters are summarized in Table~\ref{tab:setup}. Semantic distortion is modeled as i.i.d.\ Gaussian noise $\mathbf{n}\sim\mathcal{N}(\mathbf{0},\sigma_n^2\mathbf{I})$ added to the latent vector, with semantic $\text{SNR}\triangleq 10\log_{10}(P_z/\sigma_n^2)$~dB, where $P_z=\mathbb{E}[\|\mathbf{z}\|^2/D]$ is the average per-dimension latent power. During training, the \ac{snr} is uniformly sampled between \SIrange{0}{20}{\dB}; at test time, the \ac{snr} range is extended to \SIrange{-5}{30}{\dB}.

\begin{table}[!htpb]
	\centering
	\caption{Model and training configuration.}
	\label{tab:setup}
	\renewcommand{\arraystretch}{1.2}
	\begin{tabular}{@{}ll|ll@{}}
		\toprule[2px]
		\multicolumn{2}{c|}{\textbf{Model}} & \multicolumn{2}{c}{\textbf{Training}} \\
		\midrule[1px]
		\rowcolor{rowgray}
		Enc./Dec.\ layers & 6 / 6 & Optimizer & AdamW \\
		$d_\text{model}$ / $d_\text{ff}$ & 768 / 3072 & Learning rate & $10^{-4}$ \\
		\rowcolor{rowgray}
		Attention heads & 12 & Batch size & 16 \\
		Latent dim.\ $D$ & 256 & KL annealing $\beta$ & 0.01\,$\to$\,1.0 \\
		\rowcolor{rowgray}
		Max seq.\ length & 64 & Free bits $\lambda_\text{free}$ & 0.25 nats/dim \\
		Dropout & 0.1 & Word dropout & 0.5 \\
		\rowcolor{rowgray}
		Pooling & Mean & Label smoothing & 0.1 \\
		& & Early stopping & patience 15 \\
		\bottomrule[2px]
	\end{tabular}
\end{table}

Two model checkpoints are retained from the training process, at epochs~49 and~50, hereafter referred to as \emph{Codec-A} and \emph{Codec-B}, respectively. Pairing both sides with Codec-A yields an \emph{aligned} TX--RX configuration, whereas using Codec-A at the transmitter and Codec-B at the receiver creates a controlled \emph{misaligned} pair that emulates biased distortion from asynchronous model updates. To isolate the semantic-layer mechanisms, all experiments assume error-free lower-layer transmission (i.e., the \ac{phy} and \ac{mac} layers introduce no codeword errors). Performance is measured by BLEU and cosine sentence similarity, each averaged over 100~test sentences with 50~independent trials per sentence.

\subsection{Incremental Knowledge Gain}\label{sec:ik_results}

\begin{figure*}[!htpb]
	\centering
	\begin{subfigure}[b]{\linewidth}
		\centering
		\includegraphics[width=\linewidth]{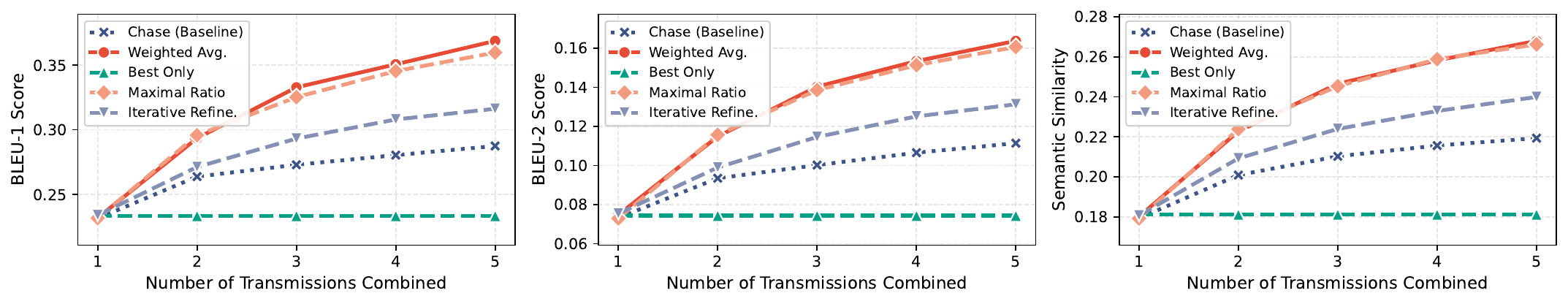}
		\caption{Aligned Codec-A/A with AWGN noise at 0~dB.}
		\label{fig:ik_awgn}
	\end{subfigure}\\
	\begin{subfigure}[b]{\linewidth}
		\centering
		\includegraphics[width=\linewidth]{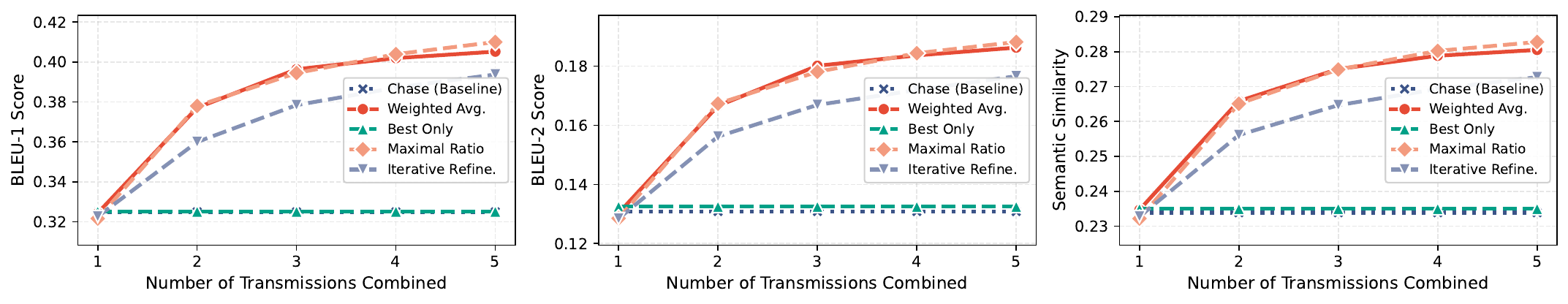}
		\caption{Misaligned Codec-A/B without additive noise.}
		\label{fig:ik_misaligned}
	\end{subfigure}
	\caption{Incremental knowledge gain vs.\ number of transmission attempts $K$ for aligned and misaligned configurations.}
\end{figure*}

We first verify that the \ac{vae}-based stochastic encoder provides meaningful \ac{ik} across retransmissions. For this purpose, we sweep the number of forced transmission attempts $K \in \{1,2,3,4,5\}$ without quality-based triggering, and compare the four combining methods from Table~\ref{tab:combining_methods} together with chase combining (where the same latent vector is used in every retransmission) as baseline. Two scenarios are evaluated: (i)~aligned Codec-A/A with AWGN-modeled semantic noise at $\text{SNR}=0$~dB, and (ii)~misaligned Codec-A/B without additive noise ($\text{SNR} \to \infty$), isolating the effect of biased distortion alone. The results are shown in Figs.~\ref{fig:ik_awgn} and~\ref{fig:ik_misaligned}, respectively. Under unbiased additive distortion alone, all methods---including chase combining---benefit from increasing $K$ due to the averaging effect, except Best-Only, whose quality scores are too uniformly low at this SNR to support meaningful selection gain. However, against biased distortion caused by model misalignment, chase combining does not provide any benefit. In both cases, the Weighted-Average and MRC-Inspired methods perform comparably and emerge as the top two methods (with Weighted-Avg.\ slightly ahead under AWGN and MRC-Inspired slightly ahead under misalignment), outperforming the Iterative method that provides an intermediate \ac{ik} gain.

\subsection{Combining Strategy Comparison}\label{sec:comb_results}

\begin{figure*}[!htpb]
	\centering
	\includegraphics[width=\linewidth]{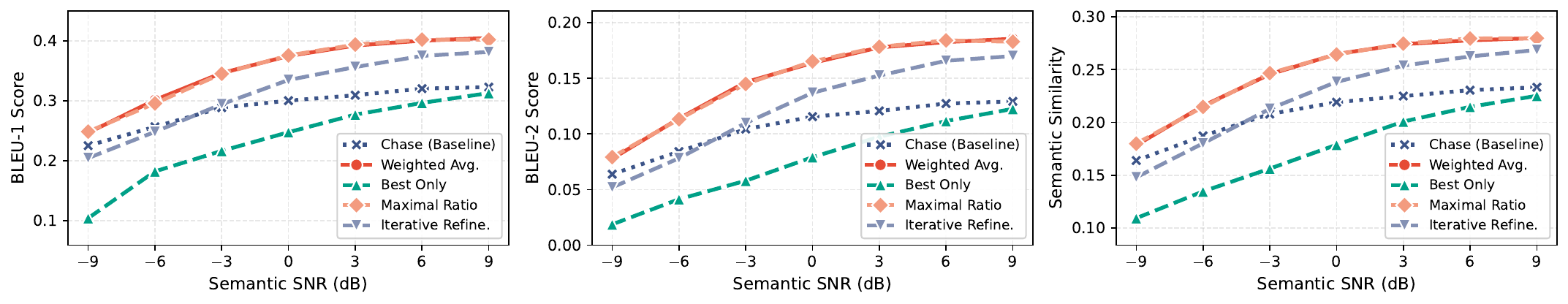}
	\caption{Comparison of different semantic soft combining methods.}
	\label{fig:combining}
\end{figure*}

Next, we benchmark the combining strategies under the joint effect of biased and unbiased distortion. The misaligned Codec-A/B pair is used with AWGN-modeled semantic noise, and the semantic \ac{snr} is swept between \SIrange{-9}{9}{\dB}. The number of transmissions is fixed to $K=5$ for all methods, bypassing quality-based triggering so that the combiners are compared under identical input conditions. The same five methods as in Section~\ref{sec:ik_results} are evaluated, and the results are shown in Fig.~\ref{fig:combining}. Consistent with the \ac{ik} gain results, the Weighted-Average and MRC-Inspired methods perform comparably as the top two, followed by the Iterative method. The Best-Only method performs poorly at low \ac{snr}, outperformed by the chase combining baseline---but the gap narrows quickly as the \ac{snr} increases.

\subsection{Quality Estimation Comparison}\label{sec:qe_results}

Finally, we evaluate the six quality estimation metrics from Table~\ref{tab:qe_methods} in a closed-loop semantic \ac{harq} setting. The combiner is fixed to Weighted-Average (Method~A), which is the best-performing strategy from Section~\ref{sec:comb_results}. The misaligned Codec-A/B configuration with AWGN semantic noise is used, with a maximum of $K_\text{max}=5$ attempts per sentence. In addition to BLEU and sentence similarity, the average number of transmissions per sentence is recorded to assess retransmission efficiency.

\begin{figure*}[!htpb]
	\centering
	\begin{subfigure}[b]{\linewidth}
		\centering
		\includegraphics[width=\linewidth]{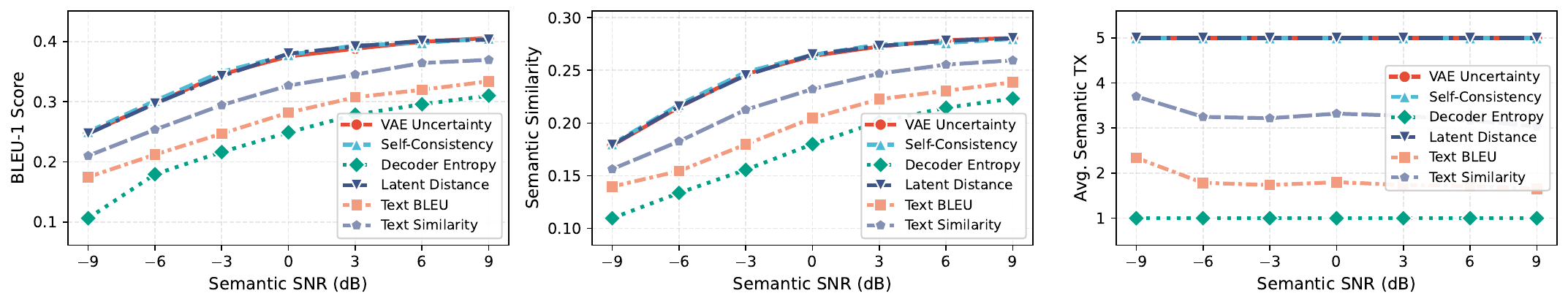}
		\caption{Sensitivity to the semantic SNR at threshold of 0.85.}
		\label{fig:quality_metrics}
	\end{subfigure}\\
	\begin{subfigure}[b]{\linewidth}
		\centering
		\includegraphics[width=\linewidth]{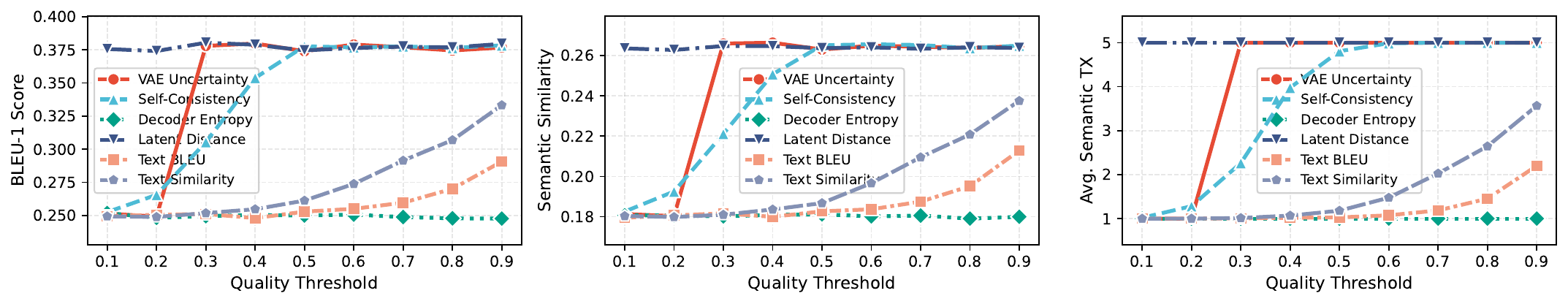}
		\caption{Sensitivity to the quality threshold at semantic SNR of \SI{0}{dB}.}
		\label{fig:qe_threshold_0db}
	\end{subfigure}
	\caption{Comparison of different semantic quality metrics.}
\end{figure*}

The evaluation proceeds in two stages. First, the quality threshold is fixed at $q_\text{th}=0.85$ and the semantic \ac{snr} is swept between \SIrange{-9}{9}{\dB} to compare the estimators' ability to detect semantic degradation across channel conditions (Fig.~\ref{fig:quality_metrics}).
Under this configuration, VAE uncertainty, self-consistency, and latent distance consistently trigger the maximum retransmissions over all \ac{snr}s; decoder entropy never triggers any retransmission; while BLEU and text similarity fall in between and respond to the \ac{snr} variation.

However, comparing estimators at a single threshold is inherently limited: the six metrics have different physical interpretations, so a fixed $q_\text{th}$ does not impose an equally stringent criterion on each. To obtain a fairer assessment, a second stage sweeps the quality threshold over $q_\text{th} \in \{0.1, 0.2, \ldots, 0.9\}$ at a representative mid-range operating point of $\text{SNR}=0$~dB (Fig.~\ref{fig:qe_threshold_0db}). This threshold sensitivity analysis reveals the differences among the metrics regarding their \emph{dynamic range} and \emph{elasticity}. While latent distance and text similarity exhibit rigid behavior at two extremes of the sensitivity, the remaining four metrics show a more gradual and smooth response to the threshold variation. Especially, the self-consistency metric demonstrates the widest dynamic range with a smooth elasticity, making it the most adaptable to different operating points and requirements.

\section{Conclusion}\label{sec:conclusion}

We proposed a \ac{harq} framework for generative semantic text communication, where a Transformer-\ac{vae} codec operates as a lightweight overlay on the conventional protocol stack. The stochastic encoder inherently provides diverse latent representations across retransmissions---enabling \ac{ik} from a single model without dedicated protocol design---while the consistent latent space permits fixed-dimensional combining with an unchanged decoder regardless of the number of retransmissions. We systematically benchmarked four quality-aware combining strategies and six semantic quality estimation methods under mixed semantic distortion. The evaluation demonstrated that our proposed framework offers significant \ac{ik} gain, especially when combining Weighted-Average or MRC-Inspired methods with self-consistency as the quality metric. Future work includes evaluating under more severe codec misalignment scenarios, extending the framework to multi-modal semantic communication, and joint \ac{harq} optimization across semantic and \ac{phy} layers.

\bibliographystyle{IEEEtran}
\bibliography{references}

\end{document}